\documentclass{aa}
\usepackage{graphicx}
\usepackage{natbib}
\usepackage{epsfig}
\def\ut#1{\mathop{\vtop{\ialign{##\crcr
     $\hfil\displaystyle{#1}\hfil$\crcr\noalign
     {\kern1pt\nointerlineskip}\hbox{$\hfil\sim\hfil$}\crcr
     \noalign{\kern1pt}}}}}

\def\undersymbol#1#2{\mathop{\vtop{\ialign{##\crcr
     $\hfil\displaystyle{#2}\hfil$\crcr\noalign
     {\kern1pt\nointerlineskip}\hbox{$\hfil#1\hfil$}\crcr
     \noalign{\kern1pt}}}}}
\def\arcsec{^{\prime\prime}}

\def\degr{^0}

\begin{document}

  \title{The X-ray eclipse of the dwarf nova HT CAS observed by the $XMM$-Newton satellite: spectral and timing analysis}
      \author{A.A.Nucita\inst{1}, B.M.T.Maiolo\inst{2}, S.Carpano\inst{3}, G.Belanger\inst{4}, D.Coia\inst{5},
              M.Guainazzi\inst{1}, F.de Paolis\inst{2,6} \and G.Ingrosso\inst{2,6}}
          \institute{XMM-Newton Science Operations Centre, ESAC, ESA, PO Box 78, 28691 Villanueva de la $\rm Ca\tilde{n}ada$, Madrid, Spain
          \and
          Dipartimento di Fisica, Universit\`a del Salento, CP 193, I-73100 Lecce, Italy
          \and
          ESA, Research and Scientific Support Department, ESTEC, PO Box 299, 2200 AG, Noordwijk, The Netherlands
          \and
          Integral Science Operations Centre, ESAC, ESA, PO Box 78, 28691 Villanueva de la $\rm Ca\tilde{n}ada$, Madrid, Spain
          \and
          Herschel Science Operations Centre, ESAC, ESA, PO Box 78, 28691 Villanueva de la $\rm Ca\tilde{n}ada$, Madrid, Spain
          \and
          INFN, Sezione di Lecce, Via Arnesano, I-73100 Lecce, Italy
          }

   \offprints{A. A. Nucita}

{
  \abstract
   {A cataclysmic variable is a binary system consisting of a white dwarf that accretes material
   from a secondary object via the Roche-lobe mechanism. In the case of long enough observation,
   a detailed temporal analysis can be performed, allowing the physical properties of the binary system to be
   determined.}
  {We present an {\it XMM}-Newton observation of the dwarf nova HT Cas acquired to resolve the binary system eclipses and constrain
   the origin of the X-rays observed. We also compare our results with previous ROSAT and ASCA data.}
   {After the spectral analysis of the three EPIC camera signals, the observed X-ray light curve was studied with well known techniques and the eclipse contact points obtained.}
   {The X-ray spectrum can be described by thermal bremsstrahlung of temperature $kT_1=6.89 \pm 0.23$ keV plus a black-body component (upper limit) with temperature $kT_2=30_{-6}^{+8}$ eV. 
Neglecting the black-body, the bolometric absorption corrected flux is $F^{\rm{Bol}}=(6.5\pm 0.1)\times10^{-12}$ erg s$^{-1}$ cm$^{-2}$, which,
   for a distance of HT Cas of $131$ pc, corresponds to a bolometric luminosity of $(1.33\pm 0.02)\times10^{31}$ erg s$^{-1}$.
   In a standard accretion scenario where $L_{BL}\simeq 0.125 L_{acc}$ assuming $\Omega _{WD} \simeq 0.5 \Omega_K(R_{WD})$, the amount of matter 
accreting onto the central white dwarf is found to be $1.7\times 10 ^{-11}$ M$_{\odot}$ yr$^{-1}$.
   The study of the eclipse in the EPIC light curve permits us to constrain the size and location of the X-ray emitting region, which turns out to be close to the white dwarf radius. We measure an X-ray eclipse somewhat smaller (but only at a level of $\simeq 1.5 \sigma$) than the corresponding optical one. If this is the case, 
we have possibly identified the signature of either high latitude emission or a layer of X-ray emitting material partially obscured by an accretion disk.
   }
   {}
}

   \authorrunning{Nucita et al.}
   \titlerunning{$XMM$-Newton observation of HT CAS}
   \maketitle
%

\section{Introduction}

Cataclysmic variables (CVs) are binary systems that consist of a white dwarf
(called primary star) gravitationally interacting
with a secondary object that is losing mass.
The two stars interact by means of the formation of a Roche-lobe,
but the details of the accretion mechanism depend upon several parameters,
the most important of which is the intensity of the magnetic field.
Depending on the intensity of the WD magnetic field, CVs are classified
as: (i) non-magnetic systems, which have a very weak field ($\ut<$0.1 MG) in which
an accretion disk forms around the compact object; (ii) intermediate polars
(with magnetic field in the range 0.1-10 MG), where the accretion disk is disrupted in the vicinity of the WD; or (iii) polars (highly magnetized systems, $\ut>$10 MG), where the accretion occurs primarily by means of a mass stream. For a detailed description of the main properties of CVs, we refer to \citet{erik}.

In non-magnetic CVs, the accretion occurs via a disk,
and according to the basic theoretical model,
half of the potential gravitational energy of the accreting material
is dissipated by the viscosity,
while the remainder is radiated away by the boundary layer (BL),
i.e., a region between the disk and the white dwarf surface.
According to this scenario, the radiation emitted by the disk reaches a maximum
in the optical and ultraviolet, while the BL, if it exists, may radiate
in the extreme ultraviolet and X-rays, with typical luminosities in the range
$10^{30}$--$10^{32}$ erg s$^{-1}$ (see e.g., \citealt{lamb82}, \citealt{baskill}).

Given the low luminosity of these systems, {\it XMM}-Newton (\citealt{jansen2001})
is remarkably well suited to their study because of its large effective area
and the possibility of observing the source simultaneously in the optical
band with the optical monitor (OM).

Interestingly, in standard accretion models for CVs, the X-ray emission is expected
to originate in the BL. Analyses of observations have demonstrated that most of the X-rays are emitted
very close to the white dwarf and that no clear correlation between the X-ray luminosity and the accretion
rate exists \citep{vantes}. Since in many CVs the observed X-ray luminosity is lower than predicted by
the boundary layer model, the details of the accretion mechanism are still unclear (but see e.g., \citealt{ferland1982} for several possible solutions to this discrepancy).

Systems with high line-of-sight inclination angles, from which eclipses are detected,
offer the unique possibility to study the location of the X-ray emitting region,
and to constrain its size when the mid ingress and egress phases of the occultation are known.
The main difficulty in this kind of study is the very low count rates, especially during an
eclipse.

An example of this kind is the well studied dwarf nova system OY Car.
In this case, the X-ray eclipse was resolved by the {\it XMM}-Newton satellite, and
\citet{ramsey2001b} demonstrated the existence of an X-ray emitting region with a
size comparable to that of the central white dwarf.
A careful analysis of the same data by \citet{oycar} showed that the
X-ray emission was probably displaced from the equatorial region and possibly related to
a magnetically controlled accretion. However, as the same authors also suggested,
the obscuration of the lower WD hemisphere by the inner accretion disk would be a valid
alternative model implying that the observed X-ray properties are caused only by projection
effects.

A CV candidate that is one of the easiest to study and constrain the size of the X-ray emitting region,
is the dwarf nova HT Cas. As for many other CVs, HT Cas is a short period binary system
($P_{orb}\simeq 1.77$ \,h, \citealt{borges}), close to the lower edge of the period gap at $\sim 2.5$\,h.
The system is characterized by a visual magnitude of $\sim 16.4$ during its quiescent state,
relatively rare outbursts (\citealt{horne}),
and white dwarf eclipses well observed in several optical campaigns.
A detailed analysis of the light curves can be found in \citet{patterson1981} and \citet{wood1985}.
In particular, \citet{wood1990} showed that the most plausible scenarios for explaining the eclipse in HT Cas correspond to an 
invisible BL or to a white dwarf totally covered by an optically thick BL.

\citet{vrielmann} were able to determine the temperature profile and surface density of the quiescent white
dwarf accretion disk with the eclipse mapping technique introduced by \citet{horne1985}.
Using the same method, \cite{feline} demonstrated the existence of variations in the disk structure,
possibly related to fluctuations in the mass transfer rate from the donor star,
and estimated the white dwarf temperature of HT Cas to be $14~000$-$15~000$ K.

There has also been a report of a cyclical change in the HT Cas orbital period
($\Delta P/P\sim 4\times 10^{-7}$, see \citealt{borges} for details). The authors attributed
this effect to a loss of angular momentum caused by the coupling of the
stellar wind from the companion star to the surface of the compact object.

We present a $\sim 45$ ks {\it XMM}-Newton observation of the dwarf nova HT Cas,
and discuss a detailed spectral and timing analysis conducted on data collected
by the EPIC cameras and the OM telescope. The paper is structured as follows:
in Sect. \ref{s:previousObs}, we briefly review the results from previous X-ray observations;
in Sect. \ref{s:obsAndResults}, we present our observation and the spectral
(Sect. 3.1) and timing (Sect. 3.2) analysis that we conducted; and
finally, in Sect. \ref{s:conclusion}, we discuss our conclusions.

\begin{figure*}[htbp]
\vspace{7.5cm} \includegraphics{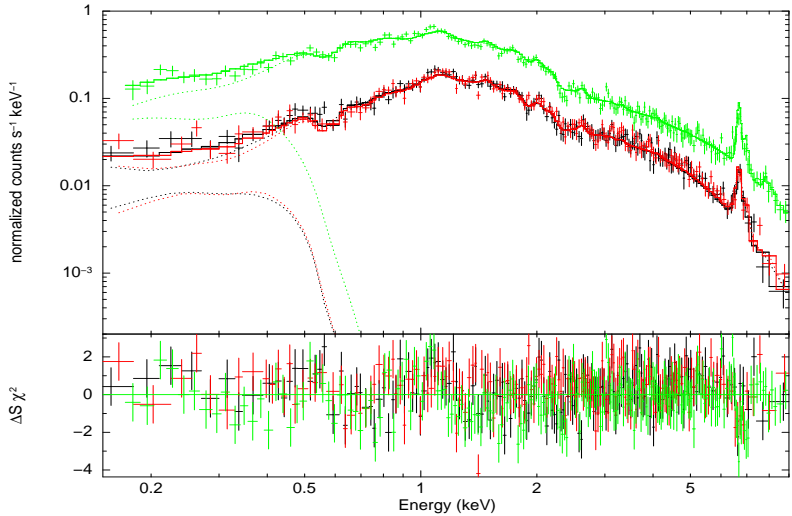}
\caption{The best-fitting model (solid lines) is given together with the MOS 1, MOS 2 and PN data
(see text for details). For clarity, we also added the individual spectral components (dashed lines) corresponding
to the MEKAL and thermal black-body models, respectively.}
\label{f1}
\end{figure*}

\section{Previous ROSAT and ASCA observation of the X-ray eclipses in HT Cas}
\label{s:previousObs}

HT Cas was observed with the ROSAT satellite for $\simeq 24$ \,h in 1991,
with an effective PSPC exposure of $5.38$\,h by \citet{woodrosat}.
Since the white dwarf temperature determined by these authors is $\sim 13~200$ K,
the detected X-rays could originate in a BL or a hot corona above and/or
below the accretion disk.
The low signal-to-noise ratio of the data did not allow the authors to determine
which of the three spectral fitting models that they used
(black-body emission, thermal bremsstrahlung and the Raymond-Smith model)
was in closest agreement with the data.
They reported a total observed flux of $\sim 5\times 10^{-13}$ erg s$^{-1}$ cm$^{-2}$,
which for the quoted distance to the target of $D\simeq 165$ pc,
corresponds to a $0.1-2$ keV luminosity of $\sim 2\times 10^{30}$ erg s$^{-1}$,
typical of other dwarf novae systems.

\citet{woodrosat} investigated the orbital variability in the X-ray light curve of HT-Cas,
discovered the existence of eclipses, and calculated the phases of mid ingress $\phi_{i}$
and mid egress $\phi_{e}$.
These points inferred an eclipse width\footnote{Note that the eclipse duration expressed in unit
of time is obtained as $\Delta T =\Delta \phi P_{orb}$, where $P_{orb}$ is the orbital period.}
at half depth of $\Delta \phi_X=0.051 \pm 0.003$ cycles. We note that this result is
consistent with those obtained using optical data, with an eclipse duration of
$\Delta \phi_{\rm opt} =0.0493 \pm 0.0007$ cycles, and ingress and egress duration of
$0.0086 \pm 0.0014$ and $0.0086 \pm 0.0012$ cycles, respectively (see \citealt{horne} for details).
Hence, they concluded that the X-ray source must be the white dwarf itself or a BL with a radius
of less than $3 R_{wd}$, since if an extended source is present it would be too faint to be detected.

Since the ROSAT observation of HT Cas was conducted during a very low flux-state,
it was impossible to carry out a detailed analysis of the X-ray light curve.
For this reason, an ASCA observation (with both SIS and both GIS cameras) during the normal quiescent state was conducted in 1994.
\citet{mukai} extracted the X-ray spectrum of  HT Cas and found that it could be reproduced well by a Raymond-Smith model (also a bremsstrahlung model) yielding a temperature close to $10$ keV, with
a column density of $n_H\simeq 3.3\times 10^{21}$ cm$^{-2}$. The observed $0.4-10$ keV flux is
$4.8\times 10^{-12}$ erg s$^{-1}$ cm$^{-2}$.

Folding the X-ray light curve at the orbital period, a clear eclipse appears
\citep[see Fig.\ 2 in][]{mukai}.
Using both a piecewise and a geometrical model \citep{horne},
an eclipse width of $0.0444_{-0.0013}^{+0.0034}$ cycles and an ingress/egress duration of
$0.0042_{-0.0034}^{+0.0041}$ cycles were derived from the ASCA X-ray light curve,
\citep[see][for details]{mukai}.

\section{{\it XMM}-Newton observation of HT Cas: data reduction}
\label{s:obsAndResults}

As seen in the previous section, analyses of both the ROSAT
and ASCA observations of HT Cas have left several open questions.
Firstly, the poor signal-to-noise ratio of the data
(particularly during the white dwarf eclipse) did not allow the investigators
to constrain the physics of the X-ray gas emitting region,
nor verify or exclude the existence of an extended source.
Secondly, the quality of the data prevented a detailed analysis of the X-ray light curve,
which then restricted the accuracy of the ingress/egress phase determination.
As \citet{mukai} suggested, an observation with as large an effective area
as that provided by the {\it XMM}-Newton satellite is required.

HT Cas was observed by {\it XMM}-Newton on two occasions in 2002
(Observation IDs 0111310101 and 0152490201), for $\simeq 45$ ks
and $\simeq 55$ ks, respectively. In this paper, we use only the first
observation, which corresponds to the normal quiescent state of HT Cas,
because the second was affected by a high level of background flaring activity for most
of the observation. The observation was conducted on $2002/08/20$ and started (ended) at
$09:23:50.0$ ($23:14:42.0$) UT.

During the first observation, HT Cas was observed by the three X-ray cameras
(RGS, EPIC MOS, and EPIC PN), and by the Optical Monitor (OM) onboard
{\it XMM}-Newton. In the present work, we ignore the RGS data given their
low statistical quality. The PN was operated in full frame mode, while the
MOS cameras were operated in large window mode, and started
$\simeq 3000$\,s after the $PN$. The OM was operated in the fast mode,
so that a light curve (in the B filter) could be obtained.

The EPIC data files (ODFs) were processed using the {\it XMM}-Science
Analysis System (SAS version $7.0.0$). The raw data were processed using the
latest available calibration constituent files. The event lists for the EPIC
cameras were obtained by running the {\it emchain} and {\it epchain} tools.

We searched for and rejected portions of the observation affected by soft proton flares,
and compiled a list of good time intervals. This resulted in exposures with effective
time of $\simeq 34$ ks, $\simeq 36$ ks, and $\simeq 19$ ks for MOS 1, MOS 2, and PN,
respectively. Only events within good time intervals were used in the spectral analysis.
The timing analysis, however, was carried out without applying good time intervals to avoid gaps, which would likely introduce spurious effects in the power spectral
analysis.

The OM data were reduced using the {\it omfchain} tool, automatically extracting
background-subtracted light curves for the sources within the fast timing window.
In the B filter ($400-500$ nm), the average out-of-eclipse magnitude is $\simeq 16.96$.


\begin{figure*}[t]
\vspace{0.2cm}
\begin{center}
$\begin{array}{c@{\hspace{0.1in}}c@{\hspace{0.1in}}c}
\epsfxsize=3.5in \epsfysize=3.1in \epsffile{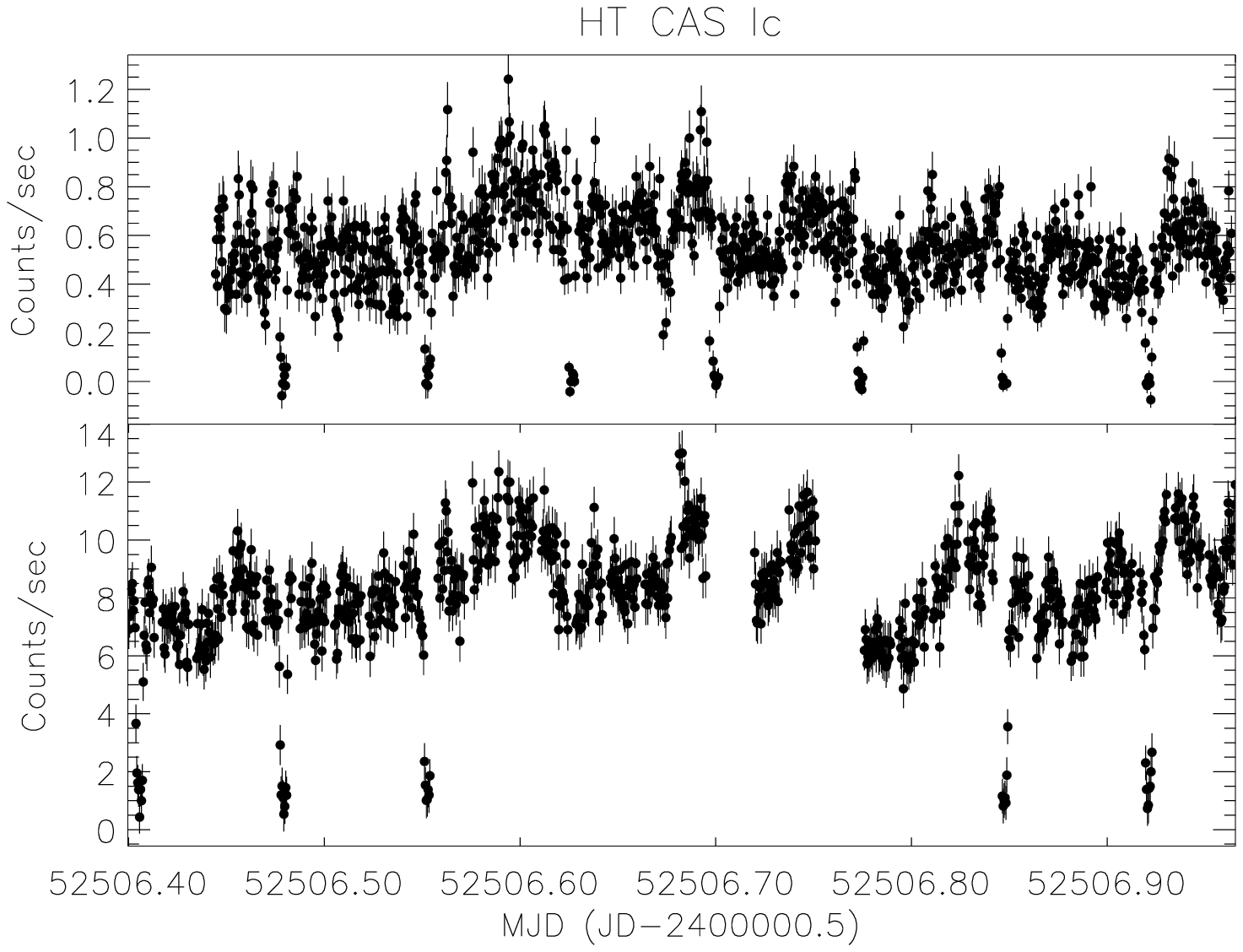} &
\epsfxsize=3.5in \epsfysize=3.1in \epsffile{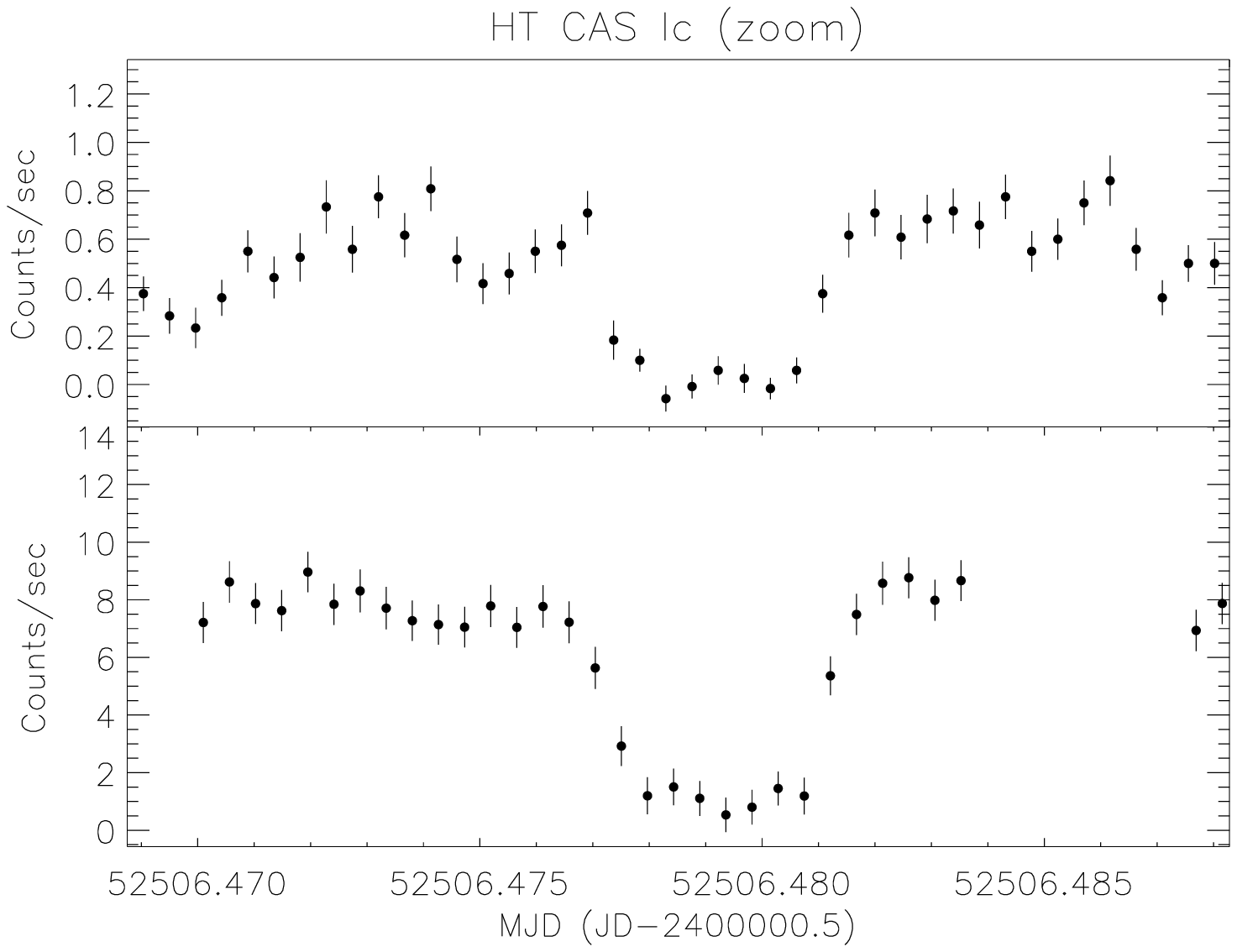}
\\
\end{array}$
\end{center}
\caption{In the upper left panel, we show the $0.2-9$ keV MOS 1, MOS 2 and PN background-corrected light curve of
the $45$ ks observation of HT Cas with a time resolution of $40$
\,s. On the bottom left panel, the OM (B filter, 400-500 nm)
background-subtracted light curve is shown with a bin size of $40$ \,s.
The large time gaps in the OM light curve correspond to ground station handovers during which the instrument did not operate.
Small time gaps are also present between subsequent OM exposures caused by overheads (see for details the \citealt{xrps}).
To the right, we show a zoom of one of the eclipses present in the data. The time axis shows barycenter-corrected time.}
\label{f2}
\end{figure*}


\subsection{Spectral analysis}
\label{s:spectral}

The EPIC source spectra were extracted in a circular region centered on the nominal
position of HT Cas in the three EPIC cameras (extraction circles with radius
$\simeq 60\arcsec$), while the background spectra were accumulated in circular
regions on the same chip and, where possible, at the same vertical location.
The net count rates of the source in the two MOS and PN cameras are
$\simeq 0.33$ count s$^{-1}$, $\simeq 0.36$ count s$^{-1}$, and $\simeq 1.00$ count s$^{-1}$.
The resulting spectra were rebinned to have at least 25 counts per energy bin,
and imported into XSPEC (version 12.4.0) for spectral fitting.

A single temperature thermal plasma model with absorption by neutral gas
(MEKAL and WABS models in XSPEC) gave a fit (with all the parameters free to vary)
with $\chi^2_{\nu}=1.3$ (549 d.o.f.), for a temperature $kT=7.14\pm 0.23$ keV and
hydrogen column density $N_H=(1.44\pm 0.04)\times 10^{21}$ cm$^{-2}$;
the normalization of this component is $N=(2.40\pm 0.02)\times 10^{-3}$.
We note that the temperature derived for this model is lower than that ($10.1\pm1.5$ keV) estimated by
using the ASCA data (\citealt{mukai}).
All the errors in this work are quoted at the $90\%$ confidence level, unless otherwise stated.

Given that this model underestimates the flux below $0.3$ keV,
a low temperature black-body component is added to obtain an acceptable fit
of $\chi^2_{\nu}=1.21$ (547 d.o.f.), yielding a temperature of $kT_1=6.89 \pm 0.23$ keV,
for the MEKAL component, and $kT_2=30_{-6}^{+8}$ eV, for the soft black-body component.
The normalizations of the two components are $N_1=(2.45\pm 0.03)\times 10^{-3}$ and
$N_2=(7_{-5}^{+20})\times 10^{-4}$, respectively. Given the very large uncertainty in the
normalisation of the black-body component, it should be considered as an upper limit.
The spectrum shows a strong emission line (iron K$\alpha$ transition)
at $6.7$ keV, which was well produced by the adopted plasma model.

The fitted hydrogen column density\footnote{%
Note that the column density obtained from the fit procedure is consistently lower than
that the average in our Galaxy ($N_H\simeq 4.1\times 10^{21}$ cm$^{-2}$, \citealt{dickey}).
}
is $N_H=(1.6\pm 0.1)\times 10^{21}$ cm$^{-2}$.
We also emphasize that the column density derived from
our fit is remarkably lower than that measured by using the ASCA data
($n_H\simeq 3.3\times 10^{21}$ cm$^{-2}$, \citealt{mukai}).
Hence, the ASCA spectrum may be affected by self-absorption at low energies that we do
not see in the {\it XMM}-Newton data. However, various column densities have been
measured, such as $1.8\times 10^{20}$ cm$^{-2}$ \citep{woodrosat}
and $6\times 10^{20}$ cm$^{-2}$ \citep{pattersona}.
As discussed by \citet{woodrosat}, in the quiescent state a fraction of the flux coming from the white dwarf, as well as from the boundary layer, may be obscured by the edge of the disk. If the obscuration does not occur during the low flux state, the column density will be lower.

In Fig. \ref{f1}, we show the MOS 1, MOS 2, and PN spectra ($0.2-9$ keV) for the source,
the irrespective best fit model (solid lines), and the individual model components
(dashed lines). The total absorbed flux in the 0.2-9 keV band is $F^{\rm{Abs}}_{\rm{0.2-9}}=(4.1\pm 0.1)\times10^{-12}$ erg s$^{-1}$ cm$^{-2}$,
the respective contributions of the MEKAL and black-body components being $F^{Mek}_{\rm{0.2-9}}=(4.05\pm 0.05)\times10^{-12}$ erg s$^{-1}$ cm$^{-2}$
and $F^{BB}_{\rm{0.2-9}}=(2.2_{-1.5}^{+6.2})\times10^{-14}$ erg s$^{-1}$ cm$^{-2}$.

We note that the measured {\it XMM}-Newton flux in the $0.1-1.2$ keV band is a factor $\sim 3$ higher than
the corresponding flux measured by \citet{woodrosat}. Therefore, our observations indicate that we are studying the normal quiescent state of HT Cas, and not a very low flux state.

Neglecting the black-body component, for a optically thin spectrum at $kT\simeq 6.89$ keV the observed
0.2-9.0 keV band flux  corrected for the absorption is $F^{\rm{Cor}}_{\rm{0.2-9}}=(5.0\pm 0.1)\times10^{-12}$ erg s$^{-1}$ cm$^{-2}$. By using Xspec, it is straightforward to estimate a bolometric correction factor of $\simeq 30\%$, therefore the unabsorbed bolometric flux turns out to be $F^{\rm{Bol}}=(6.5\pm 0.1)\times10^{-12}$ erg s$^{-1}$ cm$^{-2}$.

Consequently, for an estimated distance of $131$ pc (see \citealt{parallax}), the luminosity of HT Cas is
$(1.33\pm 0.02)\times10^{31}$ erg s$^{-1}$.

\subsection{Timing analysis}
\label{s:timing}

Light curves of the source were extracted from the original event list files to avoid introducing
gaps that can induce artifacts in the timing analysis. We applied a solar system barycentric
correction so that the event times are in barycentric dynamical time instead of spacecraft time.
For all the three EPIC cameras, we used a circular apertures of radius $\simeq 60\arcsec$ centered
on HT Cas. The background light curves were extracted from the same CCD, scaled to the extraction
area and subtracted from the source light curves. To increase the signal-to-noise ratio, the
background-corrected light curves were combined.
The X-ray (top left) and optical (bottom left) light curves of HT Cas are shown in Fig.\ \ref{f2},
with a time resolution of $40$\,s. We remind the reader that the time axis shows barycenter-corrected times.

The EPIC light curve has an average count rate of $0.56 \pm 0.05$ count s$^{-1}$, and clearly shows the presence of seven eclipses during the duration of the observation. 
As an example, a zoom in view for one of the eclipses is shown in the right panels of the same figure: EPIC (top) and OM (bottom).

These eclipses can be compared to those appearing in the bottom part of the same
figure, in which the OM (B filter) light curve is shown. We note that in the OM light curve, three eclipses
appear that are not seen because of time gaps in the observation.
In both light curves, we give the time axis in MJD (JD -$2~400~000.5$) and throughout the paper use the
linear ephemeris (barycenter-corrected) reported in \citet{borges}. 
In this respect, we use this ephemeris and note that the first eclipse appearing in the EPIC light curve occurs at the time $MJD=52~506.479~21 \pm 0.000~03$ days (as expected using the linear ephemeris).

\begin{figure*}[htbp]
\vspace{0.4cm}
\begin{center}
$\begin{array}{c@{\hspace{0.1in}}c@{\hspace{0.1in}}c}
\epsfxsize=3.2in \epsfysize=2.7in \epsffile{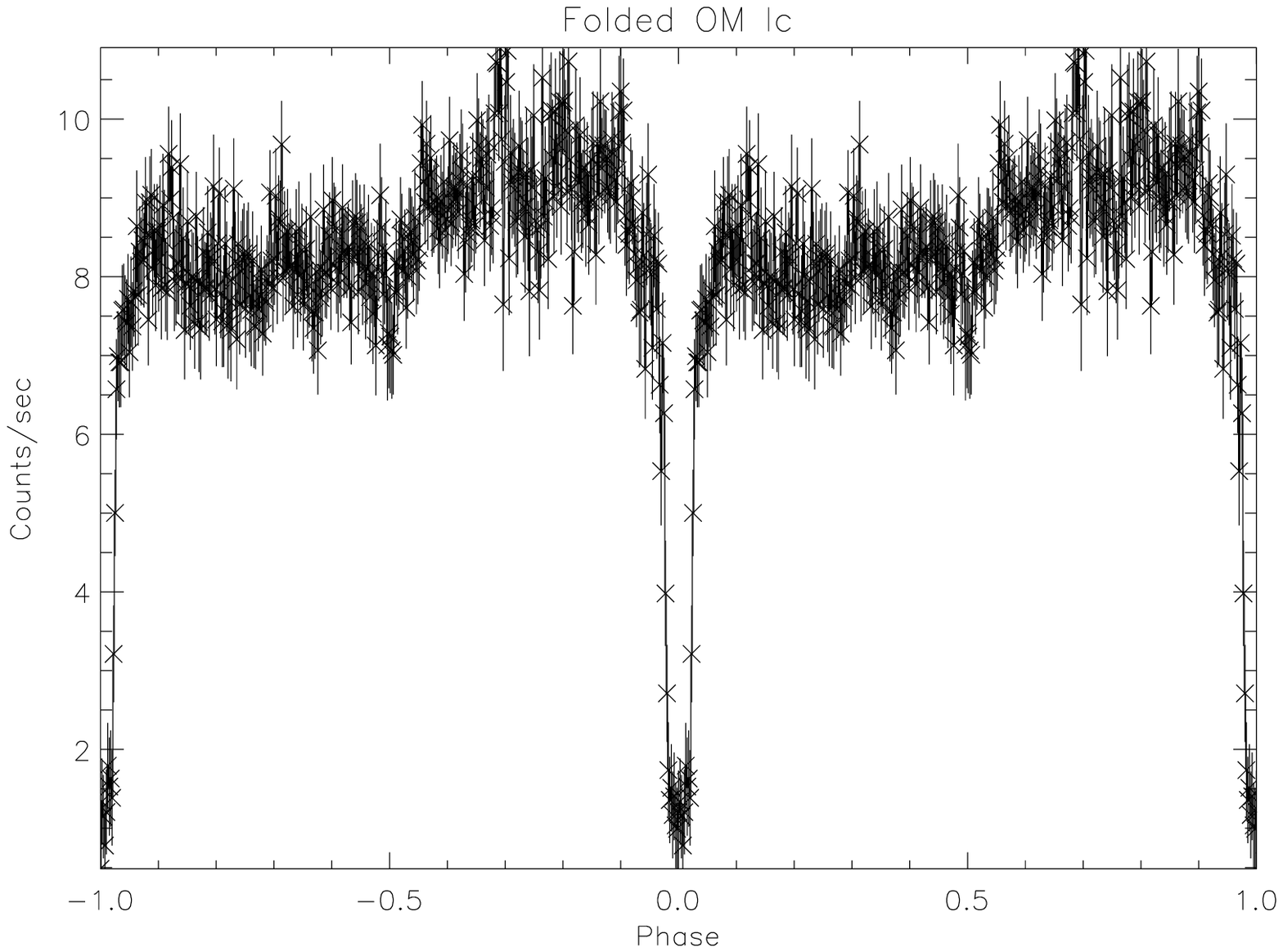} &
\epsfxsize=3.2in \epsfysize=2.7in \epsffile{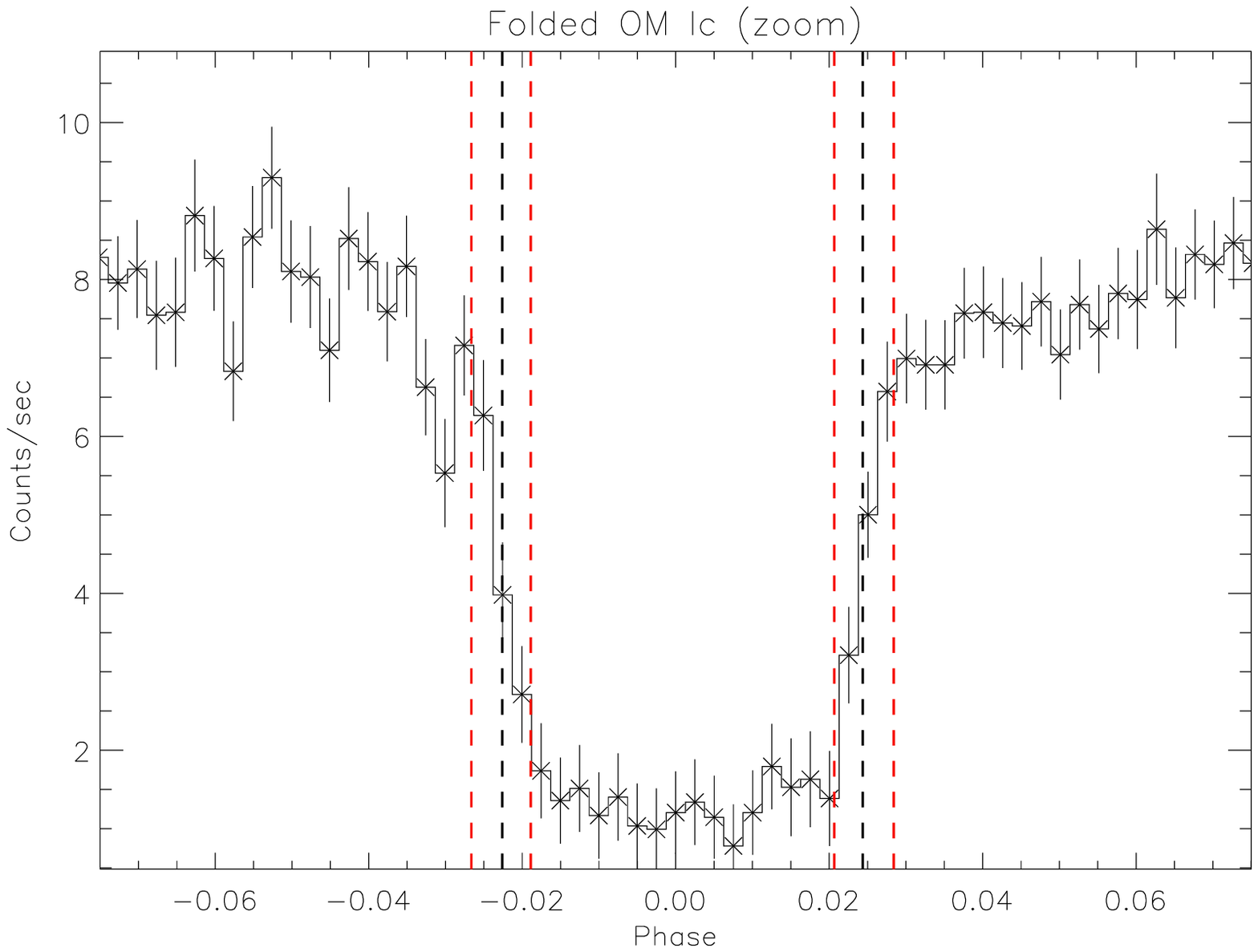}  \\
\end{array}$
\end{center}
\caption{Two-period folded OM light curve (left panel), and zoom around the HT Cas eclipse
(right panel). The red dashed vertical lines correspond to the ingress and egress contact points
for the X-ray eclipse, while the black dashed lines identify the X-ray mid ingress and egress phases.}
\label{f5}
\end{figure*}

Figure \ref{f5} presents the folded OM light curve using the
\citet{borges} linear ephemeris period of $P_0 = 0.073~647~202~9(\pm 3)$ days (the error refers to last digit),
which is equal to $\simeq 1.77$\,h \citep{borges}. The left panel shows the
two-period folded light curve, and the right shows a zoom centered on the eclipse.
We note that, in defining the phase axis of the epoch-folded light curve in Fig. \ref{f5}, we use the epoch of the first eclipse observed in the X-ray light curve as the epoch of phase 0.

\begin{figure*}[htbp]
\vspace{0.8cm}
\begin{center}
$\begin{array}{c@{\hspace{0.1in}}c@{\hspace{0.1in}}c}
\epsfxsize=3.2in \epsfysize=2.7in \epsffile{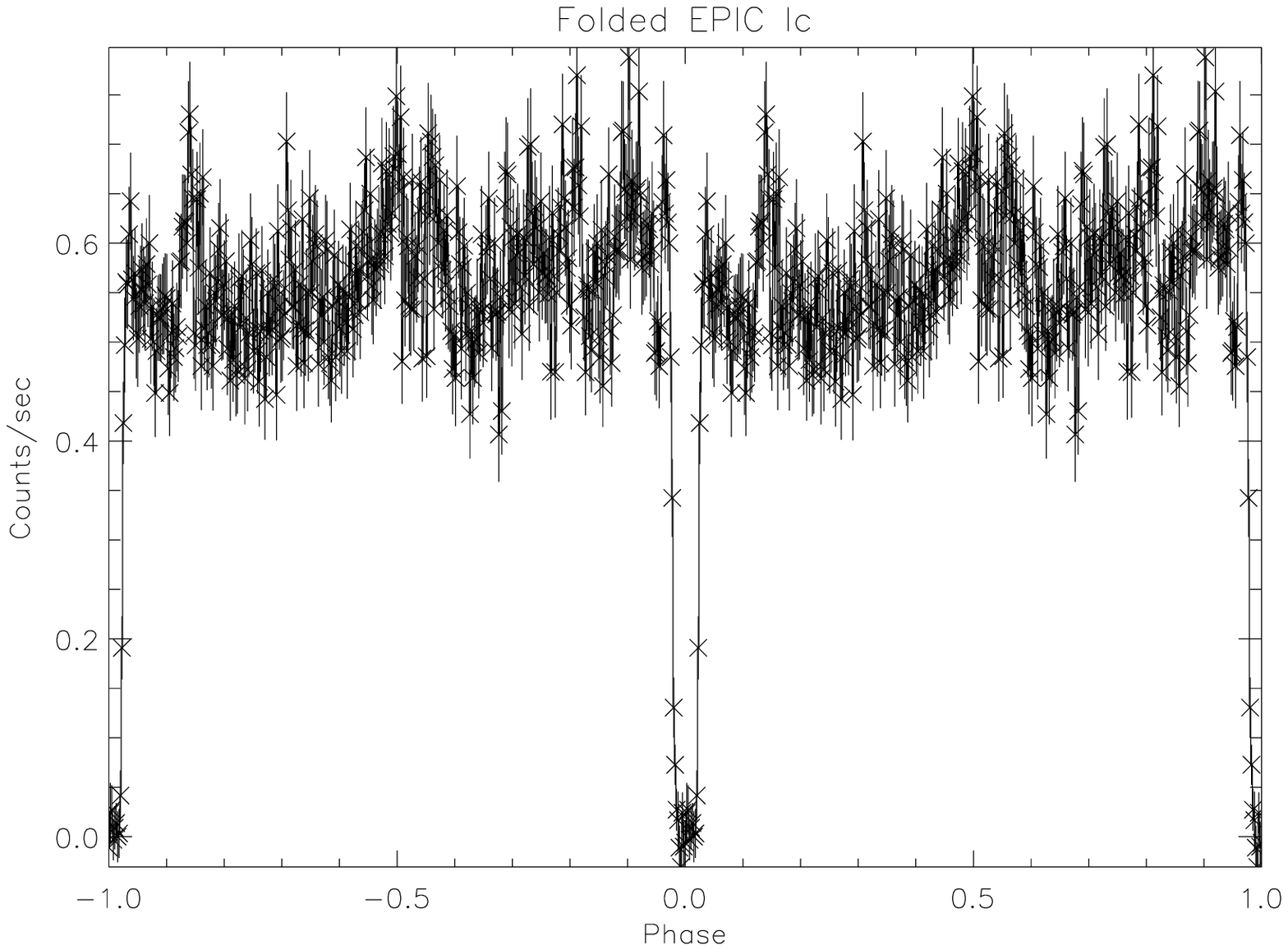} &
\epsfxsize=3.2in \epsfysize=2.7in \epsffile{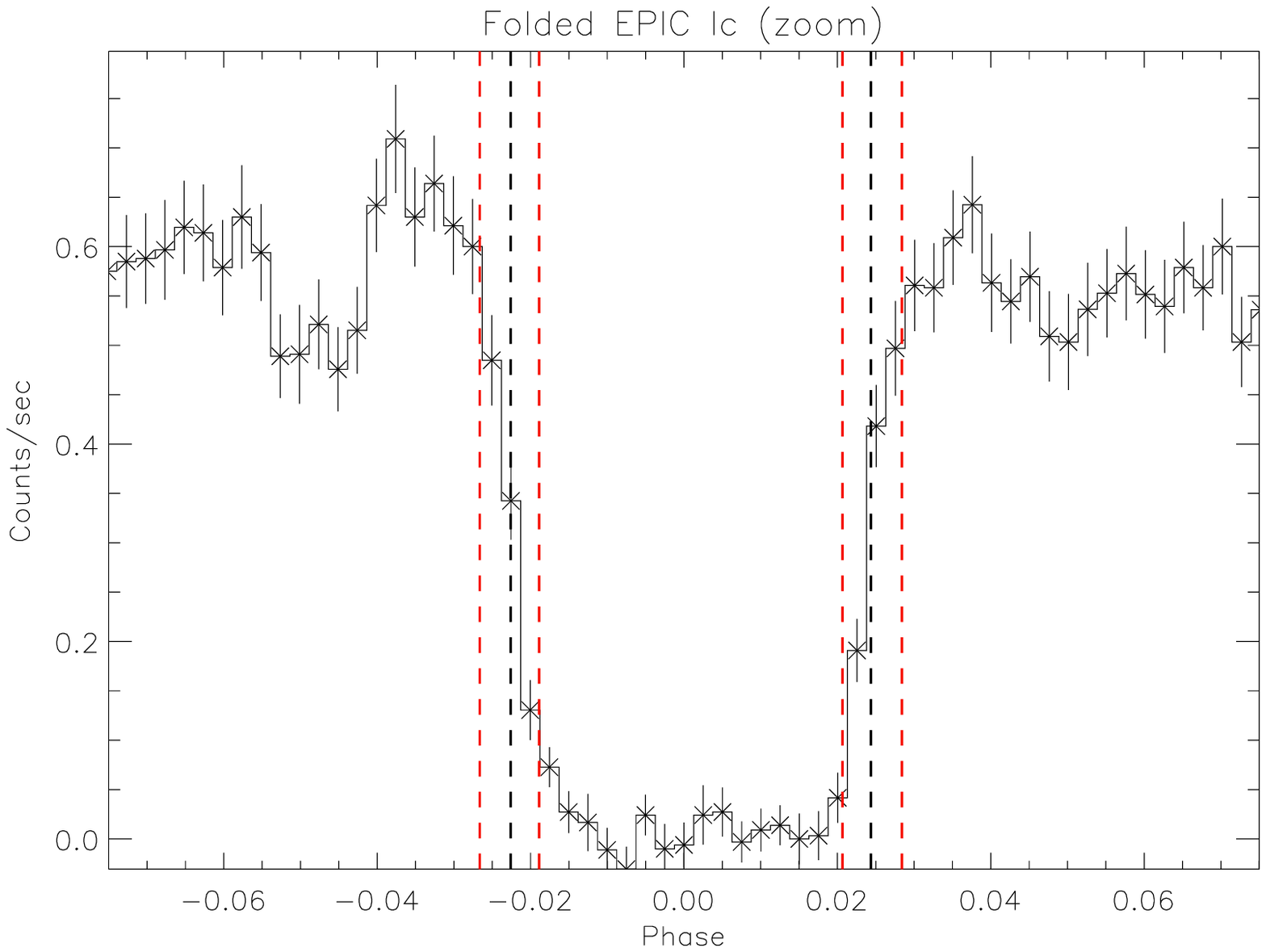}  \\
\end{array}$
\end{center}
\caption{The same as in Fig. \ref{f5} but for the X-ray light curve. See text for details about the method used to determine the eclipse contact
points.}
\label{f6}
\end{figure*}

In Fig.\ \ref{f6}, we show the EPIC light curve folded using the \citet{borges} linear ephemeris period (left panel),
and a zoom view for the eclipse (right).\footnote{%
	We applied the Lomb-Scargle technique to the X-ray light curve,
	and found a periodogram with a broad peak at 1.75\,h.
	To check the consistency of the result, we also applied the epoch-folding method to the EPIC light curve.
	This consists of folding the light curve at different periods, and fitting the result with a linear (or sine) function.
	A strong periodic signal is translated into a high (or low) $\chi^2$ value, and therefore becomes conspicuous
	in the $\chi^2$ versus period plot.
	We obtained a result consistent with the orbital period of $\simeq 1.77$\,h \citep{borges}.
}
The folded light curve was used to determine the contact points of the eclipse,
and we found that the X-ray emitting region disappears between phases
$\phi_1$ and $\phi_2$, and reappears between phases $\phi_3$ and $\phi_4$,
thus defining the duration of the ingress and egress
as  $\Delta \phi_{1,2}= |\phi_2-\phi_1|$ and $\Delta \phi_{3,4}= |\phi_4-\phi_3|$, respectively.
In addition, the mid ingress ($\phi_i$) and mid egress ($\phi_e$) points,
defined as the phases corresponding to which half of the light curve average value is eclipsed,
are used to measure the eclipse duration, i.e., $\Delta \phi_X= |\phi_e-\phi_i|$.

We proceeded in fitting the folded light curve by using a purely phenomenological
model consisting of a constant level (outside the eclipse), a flat part for the eclipse,
and two linear ingress and egress functions.
The free parameters of the piecewise function correspond to two constant levels,
the ingress $\phi_1$ and egress $\phi_4$, and the ingress/egress durations
(assumed to be identical) for a total of 5 free parameters.

\begin{figure}[htbp]
\vspace{6.5cm} \includegraphics{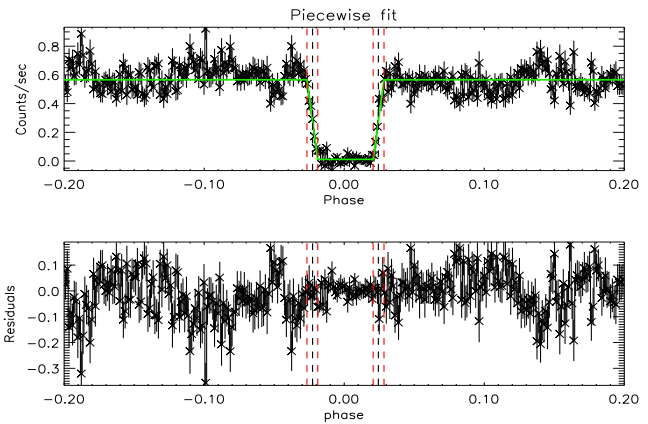}
\caption{A piecewise fit (solid green line) to the EPIC folded light curve (upper panel),
	and the comparison with the data (bottom panel).
	The dashed lines indicate the contact points derived from
	the phenomenological model (see text).}
\label{f7}
\end{figure}

Applying the Levenberg-Marquardt least squares fit to the data,
restricted around the central eclipse, yielded values of
$\phi_1= -0.0266\pm 0.0005$, $\phi_2=-0.0189\pm 0.0012$,
$\phi_3=0.0207\pm 0.0005$, and $\phi_4=0.0284\pm 0.0012$,
with a reduced $\chi^2=1.5$ (314 d.o.f.). This is illustrated in Fig.\ \ref{f7},
where the green line in the upper panel represents the overall fit,
the phases are shown by the vertical red dashed lines, and the black
dashed lines indicate the mid ingress and mid egress at
$\phi_{i}= -0.0226\pm 0.0007$ and $\phi_{e}= 0.0244\pm 0.0007$
cycles, respectively.\footnote{%
	Errors are quoted  at the $1\sigma$ confidence level,
	and are derived from the square root of the
	covariance matrix diagonal elements.
}
The values obtained for $\phi_1$, $\phi_2$, $\phi_3$, and $\phi_4$ imply
ingress/egress and mid eclipse durations of $\Delta \phi_{1,2} = \Delta \phi_{3,4} = 0.0077 \pm 0.0017$ and $\Delta \phi_X=0.0469 \pm 0.0014$
cycles, respectively. In the time domain, the ingress/egress and total occultation time
scales of the X-ray light curve of HT Cas are $49.2\pm 9.9$ \,s, and $298.9\pm 8.9$ \,s,
respectively. For clarity, we also give the results of the piecewise fit to the data in Table 1.

The duration of the eclipse in our X-ray light curve is consistent
with the white dwarf eclipse observed in the ROSAT and ASCA
data \citep{woodrosat,mukai}. As in the ASCA data, the eclipse duration that we measure seems to be shorter
(but only at the low significance level of $\sim 1.5 \sigma$) with
respect to the optical observation \citep[$\Delta \phi_{\rm opt}=0.0493\pm0.0007$ cycles, ][]{horne}.
In addition, for the first time, we have an accurate measure of the ingress and egress
durations, the length of which are consistent with the optical counterpart
\citep[$\Delta \phi_{\rm 1,2,opt}=0.0086 \pm 0.0014$ cycles, and $\Delta \phi_{\rm 3,4,opt}=0.0086 \pm 0.0012$ cycles,][]{horne}. This point is discussed at length in Sect. 4.
Applying the same analysis to the folded OM light curve (B filter) yields results
that are consistent with those found in \citet{horne}.


The eclipse depth in the EPIC light curve of HT Cas
was measured, following the procedures described by \citet{mukai}, by
determining the background-corrected count rates inside and outside
the eclipse phases after either 1) extracting the spectra in the corresponding time intervals,
or 2) directly from the folded EPIC light curve.
The two methods provided results that were equivalent within the uncertainties:
0.01 $\pm$ 0.02 count s$^{-1}$ during the full eclipse duration
($\phi_2$ to $\phi_3$), and 0.56 $\pm$ 0.05 count s$^{-1}$ outside.
The results are similar in the soft and hard bands, as found by \citet{mukai}.

\begin{table}
\begin{center}
\begin{tabular}{|l|c|}
\hline
{\bf Quantity} & {\bf Value}  \\
\hline
First X-ray eclipse (day)           & MJD (JD -$2~400~000.5$)\\
       &   $52~506.479~21\pm 0.000~03$  \\
\hline
Ingress phase $\phi_1$ (cycles) & $-0.0266\pm 0.0005$ \\
Ingress phase $\phi_2$ (cycles) & $-0.0189\pm 0.0012$ \\
Egress phase $\phi_3$ (cycles) & $ 0.0207\pm 0.0005$ \\
Egress phase $\phi_4$ (cycles) & $0.0284\pm 0.0012$ \\
                            &                    \\
Ingress $\Delta \phi_{1,2}$ (cycles) & $0.0077\pm 0.0017$ \\
Egress  $\Delta \phi_{3,4}$ (cycles) & $0.0077\pm 0.0017$ \\
                            &                    \\
Ingress $\Delta t_{1,2}$ (s) & $49.2\pm 9.9$ \\
Egress  $\Delta t_{3,4}$ (s) & $49.2\pm 9.9$ \\
\hline
Mid ingress $\phi_i$ (cycles) & $-0.0226\pm 0.0007$ \\
Mid egress $\phi_e$ (cycles) & $0.0244\pm 0.0007$ \\
                            &                    \\
Eclipse $\Delta \phi_X$ (cycles) & $0.0469\pm 0.0014$ \\
                            &                    \\
Eclipse $\Delta t_X$ (s) & $298.9\pm 8.9$ \\
\hline
Mid-eclipse rate (counts s$^{-1}$) &  $0.01 \pm 0.02$   \\
Out-of-eclipse rate (counts s$^{-1}$) &  $0.56 \pm 0.05$ \\
\hline
\end{tabular}
\caption{{\it XMM}-Newton X-ray eclipse of HT-Cas: result of the piecewise fit (see text for details).}
\end{center}
\label{table1}
\end{table}

\section{Results and discussion}
\label{s:conclusion}

In this paper, we have focused on the {\it XMM}-Newton observation of the eclipsing dwarf nova HT Cas.
The source was observed for $\simeq 45$\,ks, and thus a good quality spectrum
and light curve can be obtained from the MOS and PN cameras.

The EPIC spectra can be fitted simultaneously using XSPEC
with an absorbed thermal plasma model, to which the addition of a black-body component improves the fit.
The best-fit model ($\chi^2_{\nu}=1.21$, for 547 d.o.f.) yields parameter values
of $kT_1=6.89 \pm 0.23$ keV, for the MEKAL temperature, and $kT_2=30_{-6}^{+8}$ eV,
for the soft black-body.
The hydrogen column density is found to be $N_H=(1.6\pm 0.1)\times 10^{21}$ cm$^{-2}$,
and the normalizations of the two components are
$N_1=(2.45\pm 0.03)\times 10^{-3}$ and $N_2=(7_{-5}^{+20})\times 10^{-4}$, respectively.

Since the black-body flux in the X-ray band may be considered to be an upper limit, in the following discussion it is not considered. Hence, for a optically thin spectrum at $kT\simeq 6.89$ keV, the observed
0.2-9.0 keV band flux is $F^{Mek}_{\rm{0.2-9}}=(4.05\pm 0.05)\times10^{-12}$ erg s$^{-1}$ cm$^{-2}$, which corresponds to an absorption corrected flux of $F^{\rm{Cor}}_{\rm{0.2-9}}=(5.0\pm 0.1)\times10^{-12}$ erg s$^{-1}$ cm$^{-2}$. By using Xspec, it is possible to evaluate a bolometric correction factor of the order $\simeq 30\%$, so that the bolometric flux turns out to be $F^{\rm{Bol}}=(6.5\pm 0.1)\times10^{-12}$ erg s$^{-1}$ cm$^{-2}$. For an estimated distance to HT Cas of 131 pc, the bolometric luminosity comes out to be
$(1.33\pm 0.02)\times10^{31}$ erg s$^{-1}$.

As demonstrated by \cite{popham1995}, the total luminosity of the accretion disk
of a white dwarf of mass $M_{WD}$ and radius $R_{WD}$ is $L_{disk}=GM_{WD}\dot{M}/2R_{WD}$.
Thus, only half of the gravitational potential energy $L_{acc}=GM_{WD}\dot{M}/R_{WD}$
is radiated away by the disk, while the other half is in the form of rotational energy in the
accreting material. A fraction of this energy is dissipated at the boundary layer,
where the velocity of the accreting gas decreases, with a rate depending on the rotational
speed of the white dwarf. Taking into account the white dwarf spin, the boundary layer luminosity implies that $L_{BL}\simeq 0.125 L_{acc}$ for
$\Omega _{WD} \simeq 0.5 \Omega_K(R_{WD})$ \citep{popham1995}.
Since the respective best-fit model white dwarf mass and radius of 0.61 M$_{\odot}$ and
0.0118 R$_{\odot}$ are derived from the photometric measurements \citep{horne},
one can use the standard accretion scenario (and the bolometric luminosity quoted above) to estimate the mass accretion rate, which, for HT Cas, turns out to be $1.1\times 10^{15}$ g s$^{-1}$ or, equivalently,
$1.7\times 10^{-11}$ M$_{\odot}$ yr$^{-1}$ assuming $\Omega _{WD} \simeq 0.5 \Omega_K(R_{WD})$.
Of course, since we have neglected the luminosity of the black-body component, the previous estimate for the
mass accretion rate must be considered to be a lower limit.


\begin{figure}[htbp]
\vspace{8.5cm} \includegraphics{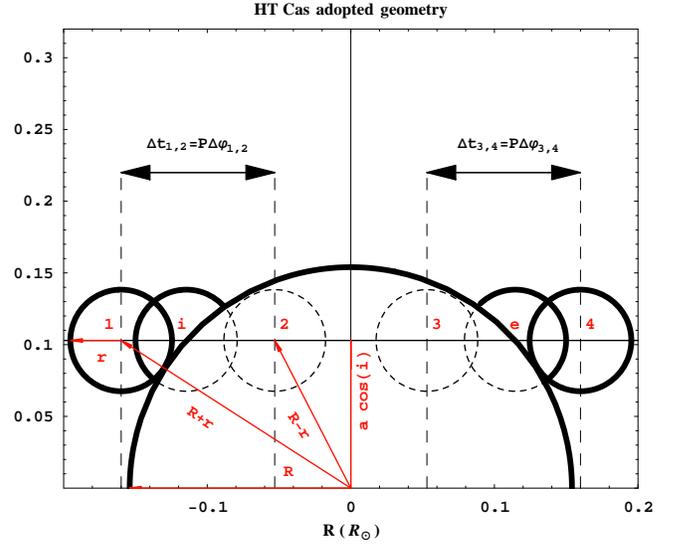}
\caption{The adopted geometry for the eclipse description in HT Cas (see text for details).}
\label{f8}
\end{figure}

The accurate determination of the characteristics of the eclipse based upon the
analysis of the folded EPIC light curve (Table 1), can now be used to constrain the size and
location of the X-ray emitting region.
Figure \ref{f8} presents the adopted geometry, closely following the model of \citet{robinson}:
the binary system, consisting of the central white dwarf (with radius $r$) and its companion star (with radius $R$), is viewed orthogonal to the line of sight.
The circles drawn in Fig. \ref{f8} represent the limbs of the companion star and white dwarf, whose radii ($R\simeq 0.154$ R$_{\odot}$ and $r\simeq 0.0118$ R$_{\odot}$, respectively; see \citealt{horne}) 
are determined by optical observation. Here, we have increased the radius $r$ of the white dwarf for clarity. Consequently, the two stars are separated by a distance of $a \cos i$, where  $a\simeq 0.658$ R$_{\odot}$ and $i\simeq 81^{\degr}$ are the orbital separation and the inclination angle, respectively \citep[see][for more details]{horne}.

Assuming that the cross sections of both the source ($r$) and the obscuring
star ($R$) remain constant over the eclipse duration,
the source disappears behind the companion limb between the positions labeled as $1$ and $2$,
and reappears between the positions $3$ and $4$, while the positions
labeled as $i$ and $e$ correspond to the mid-ingress and mid-egress points, respectively.
The timescales for the eclipse ingress/egress ($\Delta t_{1,2}$/$\Delta t_{3,4}$) are also reported in Fig. \ref{f8}.

It is therefore straightforward to derive the relation between the ingress $\Delta t_{1,2}$ (or egress $\Delta t_{3,4}$) timescale and the orbital parameters (period $P$, separation $a$, inclination $i$) as well as the sizes of the emitting region $r$ and eclipsing object $R$, i.e., (see also \citealt{ww2002})

\begin{equation}
\begin{array}{l}
\Delta t_{1,2}=\frac{P}{2\pi} \left[\sqrt{ \left( \frac{R+r}{a}\right)^2 -\cos ^2 i} - \sqrt{ \left( \frac{R-r}{a}\right)^2 -\cos ^2 i }\right].\\
\end{array}
\label{eq3}
\end{equation}

Using the measured X-ray ingress/egress ($\Delta t_{1,2}/\Delta t_{3,4}$) timescale and the orbital parameters determined by the optical observation (i.e., $P$, $a$, $i$, and $R$ in \citealt{horne}), 
from Eq. (\ref{eq3}) one can easily derive the size\footnote{Of course, the model outlined here is an over-simplification since we are considering a circular orbit and a constant size for the projected source.} 
of the X-ray emitting region of $r_X = (0.0117\pm 0.0004)$ R$_{\odot}$.

When this result is compared to the size of the white dwarf ($r\simeq 0.0118$ R$_{\odot}$)
of \citet{horne}, it clearly indicates that the size of the X-ray emitting region is comparable to
that of the white dwarf, which allows us to tightly constrain the origin of X-rays in HT Cas.

However, the piecewise linear fit to the EPIC light curve infers an X-ray eclipse width of
$\Delta \phi_X =0.0469\pm 0.0014$ cycles, which seems to be narrower (but, as stated in Sect. \ref{s:timing}, at the low significance level of $1.5\sigma$) 
than in the optical band \citep[$\Delta \phi_{\rm opt}=0.0493 \pm 0.0007$][]{horne}, as also reported by \citet{mukai}.

If one really believes that an X-ray eclipse is slightly narrower than in the optical, a possible explanation could be 
that the X-rays originate in a layer very close to the surface of the white dwarf, whose southern hemisphere is obscured by material that is optically thick to X-rays, 
while allowing the optical emission to reach the observer.

A different scenario could also be depicted. Given that the eclipse width is measured from the mid-ingress ($i$) to the mid-egress ($e$) points (take a closer look at Fig. \ref{f8}), 
a shorter eclipse duration implies a shorter crossing distance. Therefore, we can infer the position of the X-ray source along
the binary axis orthogonal to the orbital plane.
If the optical eclipse corresponds to the occultation of the white dwarf,
the narrowness of the X-ray eclipse suggests that the emitting region
is located above the orbital plane. If we now assume that this region
coincides with the pole of the white dwarf, it is easy to see that the
start of the optical eclipse precedes that of the X-ray eclipse,
while both end simultaneously.
The same scenario is also naturally applicable to the egress events, but in the opposite sense.

We note that a similar geometry was used by \citet{oycar} to conclude that the X-ray
emission in the dwarf nova OY Car could originate in the polar region of the
central white dwarf (see e.g. their Fig. 5), and to show how magnetic
accretion may play an important role in this kind of system. However, as the same authors also suggested,
the obscuration of the lower WD hemisphere by the inner accretion disk would be a valid
alternative model (possibly more reasonable) implying that the observed X-ray properties are caused only by projection effects.

When comparing the X-ray ingress/egress phases in HT Cas (Table 1)
with those derived in optical \citep[see Table 5 in][]{horne}, it is clear that the
geometry described above may be applied, and polar accretion may be invoked
to explain our X-ray light curve (keeping measurement uncertainties in mind). In this respect, we also note that the OM data provides a description of the optical 
light curve consistent with that found in the literature but, of course, deeper and longer observation possibly covering more cycles are required to differentiate between the two scenarios described above.

\begin{acknowledgements}
We are grateful to the referee, Julian P. Osborne, for spotting several crucial
points in the first version of the paper and for detailed reports that
significantly improved the manuscript. This paper is based on observations from XMM-Newton, an
ESA science mission with instruments and contributions directly funded by
ESA Member States and NASA.
B.M.T. Maiolo acknowledges support from the Faculty of the European Space Astronomy Centre (ESAC)
and A.A.N. is grateful to the Observatory Cassini of Loiano (Bologna, Italy) for the kind hospitality.
We are grateful to Erik Kuulkers for the interesting discussions.
\end{acknowledgements}


\end{document}